# Designing Materials Acceleration Platforms for Heterogeneous $CO_2$ Photo(thermal)catalysis

Andrew Wang [a, b], Carlota Bozal-Ginesta [a, b], Sai Govind Hari Kumar [a, b], Alán Aspuru-Guzik [a, b], Geoffrey A. Ozin*[a]

[a] Department of Chemistry, University of Toronto, Lash Miller Chemical Laboratories, 80 St George Street, Toronto, ON M5S 3H6

[b] Department of Computer Science, University of Toronto, Sandford Fleming Building, 10 King's College Road, Toronto, ON M5S 3G4

* Corresponding author: g.ozin@utoronto.ca

**Abstract**

Materials acceleration platforms (MAPs) combine automation and artificial intelligence to accelerate the discovery of molecules and materials. They have potential to play a role in addressing complex societal problems such as climate change. Solar chemicals and fuels generation via heterogeneous $CO_2$ photo(thermal)catalysis is a relatively unexplored process that holds potential for contributing towards an environmentally and economically sustainable future, and therefore a very promising application for MAP science and engineering. Here, we present a brief overview of how design and innovation in heterogeneous $CO_2$ photo(thermal)catalysis, from materials discovery to engineering and scale-up, could benefit from MAPs. We discuss relevant design and performance descriptors and the level of automation of state-of-the-art experimental techniques, and we review examples of artificial intelligence in data analysis. Based on these precedents, we finally propose a MAP outline for autonomous and accelerated discoveries in the emerging field of solar chemicals and fuels sourced from $CO_2$ photo(thermal)catalysis.

**Introduction**

Among the various methods for storing solar energy and recycling carbon dioxide ($CO_2$) to value-added chemicals and fuels, gas-phase $CO_2$ heterogeneous photo(thermal)catalysis[1] has been relatively unexplored when compared to the variety of approaches currently investigated, such as electrocatalysis, biocatalysis, thermocatalysis, solar thermocatalysis, and photocatalysis. The process combines photochemistry and photothermal chemistry to induce the reduction of gaseous $CO_2$ on the surface of a solid catalyst, offering a technologically simpler process and potentially lower energy costs.[2,3] It can be implemented by solar or light emitting diode powered gas-phase flow systems that can be scaled, lowering the barrier for finding alternatives to traditional energy and carbon intensive thermochemical reactors that form the basis of the modern catalytic industry. Yet, photo(thermal)catalysis is complex and much science still needs to be understood, which hampers the discovery of new catalyst materials and hinders the optimization of those that exist. Beyond that, a scaled-up version of a photocatalyst in commercial operation will likely exist in quite different form than that discovered in the laboratory,[4] and the challenge of how to best to utilize the photocatalyst in a larger-scale photoreactor environment needs to be urgently addressed. Because of the manifold variables affecting gas-phase heterogeneous $CO_2$ photo(thermal)catalysis, we believe this is an area that is poised to be advanced by materials acceleration platforms (MAPs), defined as closed-loop self-driving systems that incorporate high-throughput automated experimentation and analysis, and "learn" from the data with artificial intelligence (AI) and machine learning (ML).[5] Herein, we identify the physicochemical descriptors and techniques that should be considered to develop gas-phase heterogeneous $CO_2$ photo(thermal)catalysis, and we discuss their potential integration into a MAP.

Photo(thermal)catalysis refers to both photocatalysis and photothermal catalysis, the latter of which can occur concurrently and in the same environment as the former, either unintentionally or induced intentionally, depending on the catalytic material, light intensity and wavelength used, operating temperature, non-radiative electron-hole relaxation pathways, and design of the photocatalyst-photoreactor combination. The conversion of $CO_2$ in the gas phase via solar thermochemistry, which uses heat from concentrated solar power to drive reactions, has been well-established over the past decade.[6–9] A working catalyst involving a solid oxide can, under sufficient input energy, be made to form oxygen vacancies, thereby releasing oxygen gas and becoming non-stoichiometric. The non-stoichiometric form of these oxides then interact with small oxygen-containing molecules such as $CO_2$ and abstract their oxygen atoms to refill the oxygen vacancies generated earlier in the catalyst. In solar thermochemistry, this thermodynamically-uphill and kinetically-demanding process is extremely energy intensive, typically operating in the 1000–1500 °C temperature range. Alternatively, part of the required energy can be provided by the absorption of light and co-production of heat, following a photo(thermal)catalytic approach. Recent studies have demonstrated proof-of-concept photo(thermal)catalysis using nanostructured $In_2O_3$,[10–13] Cu-doped $TiO_2$,[2] and a variety of metal oxides and oxychlorides.[3] When ultraviolet-visible (UV-vis) radiation is directly involved, the same reaction can occur at temperatures around 100 – 200 °C enabled by photo(thermal)catalysis,[2,3] demonstrating that the direct interaction of light with the catalyst can help advance the reaction coordinate through specific excited states. It can be difficult however to completely decouple the thermochemical and photochemical contributions to the processes, and the detailed mechanistic pathways and the material structural-property-performance relationships remain somewhat of a mystery.

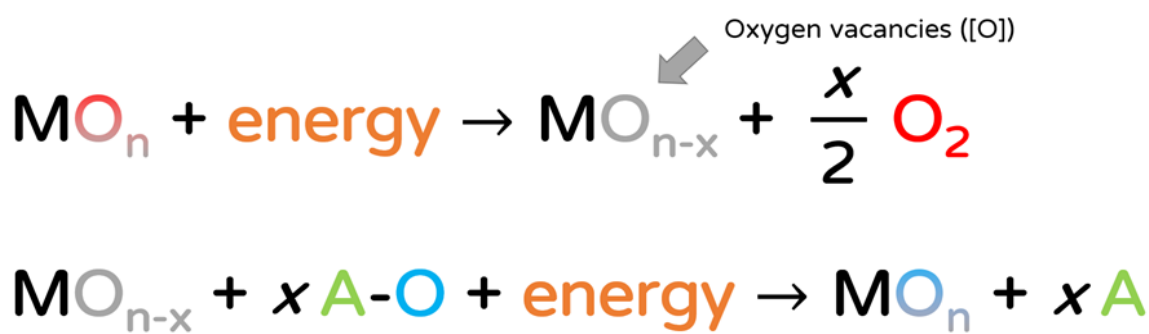

**Scheme 1.** A gas-phase heterogeneous catalytic cycle between an oxygen-containing molecule A-O (e.g., $H_2O$, $CO_2$) and a metal oxide (MO), made non-stoichiometric under input energy. In thermochemistry, this energy (directly involved in the reaction) is heat. The challenge for photo(thermal)catalysis is to combine light and co-generated heat as energy input under much milder experimental conditions.

For the above mechanism, the photo(thermal)catalytic route is relatively underdeveloped compared to that seen for thermochemical redox cycling. Examples in the literature on photo(thermal)catalysis via this pathway are rare and the efficiencies are low (typically around 1%).[2,14] However, we believe it holds significant potential, especially when considering the possibilities offered by photothermal chemistry[15] and a recent example of a record-high quantum yield seen for the photoreforming of methanol.[16]

Industrially relevant $CO_2$ conversion to $C_{1+}$ hydrocarbons via any flavour of photocatalysis would require a massive increase in production rates, efficiencies, and catalyst stabilities compared to what is currently available to bring the final cost per kg ($/kg) of product comparable to current fossil-based methods

(for example, industrial steam methane reforming to syngas currently operates at 20,000 to 200,000 normal cubic meters per hour (Nm$^3$/h);[17] if assuming a H$_2$/CO ratio of 1:2 (which gives density = 0.84 kg/m$^3$, molar mass = 19 g/mol), this translates to the order of $10^6 - 10^7$ mol/h, which even for a catalyst with a currently exceptional 1 mol (g catalyst)$^{-1}$·h$^{-1}$ requires 1 – 10 metric tons to achieve – this number can only be brought down by increasing the catalyst turnover; on top of this, stable catalysts operating at > 10% solar-to-fuel efficiencies would also be needed[18]). As a result, there is significant opportunity space for advancing CO$_2$ photo(thermal)catalysis with MAPs.

The challenge to develop gas-phase heterogeneous CO$_2$ photo(thermal)catalysis is two-fold.[19] First, there is a scientific challenge to understand how the material composition and structure influences the catalytic reaction and performance, and to come up with the best catalyst material possible. Second, there is an engineering task on how to best use and scale up the process by designing the optimal photocatalyst-photoreactor combination and reaction conditions. Addressing these aspects is necessary to achieve a viable industrial CO$_2$ photocatalytic process. Therefore, when developing a MAP for gas-phase heterogeneous CO$_2$ photo(thermal)catalysis, as in manual empirical research, it is necessary to measure or calculate descriptors representing appropriate aspects of both materials and reactors. A few relevant universal descriptors are collected in Scheme 2, where they have been split into two main groups, design descriptors and performance descriptors (e.g. the features and targets, respectively, in training a ML model) both relating to the photocatalyst and photoreactor. The goal of a MAP would be to arrive at an optimal set of performance descriptors as an outcome of an AI-driven optimization process over several series of experiments.[20,21]

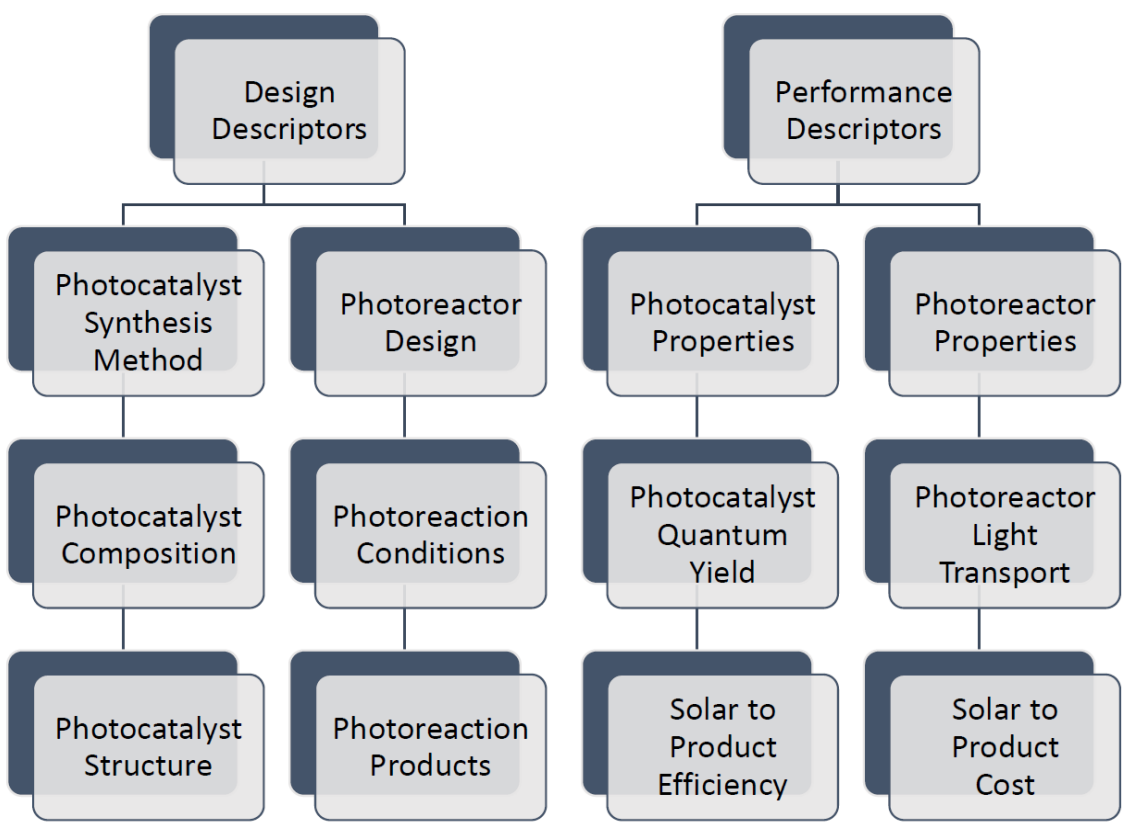

**Scheme 2**. Design and performance descriptors (e.g. features and targets) for optimizing gas-phase CO$_2$ heterogeneous photo(thermal)catalysis via a MAP.

The measurement of relevant performance descriptors could be integrated into an automated workflow for gas-phase CO$_2$ heterogeneous photo(thermal)catalysis to systematically generate experimental datasets. Based on the descriptors above, we next discuss the automation level of the corresponding techniques. Finally, we move on to propose how machine learning could be interfaced within the associated MAP to process the experimental and theoretical datasets effectively.

**Automating materials science and photo(thermal)catalysis**

The preparation and characterization of photo(thermal) catalytic materials and reactors, and their corresponding descriptors, depend on a range of common experimental techniques. To assess the potential to automate and synchronize these techniques, we first evaluate their automation level in Scheme 3 following four different criteria. The first criterion is related to the architecture of the experimental setup and to the number of different experimental techniques it can execute. An experimental setup may consist of (1) a centralized infrastructure with limited access and few available techniques, (2) commercial standalone modules, or (3) modules synchronized by a control system and / or robotic aids. This criterion may be related to the cost and degree of specialization required to automate a given technique. The second and third criteria are based on the automation level of the measurement control and analysis, and on the high-throughput sample capacity, respectively. They affect the amount of human time required to run multiple samples and to extract representative parameters. The last criterion depends on the use of machine learning to support the data interpretation and the decision-making processes, which typically relies on the availability of relevant sets of data. Each one of these criteria should be considered individually when developing a MAP.

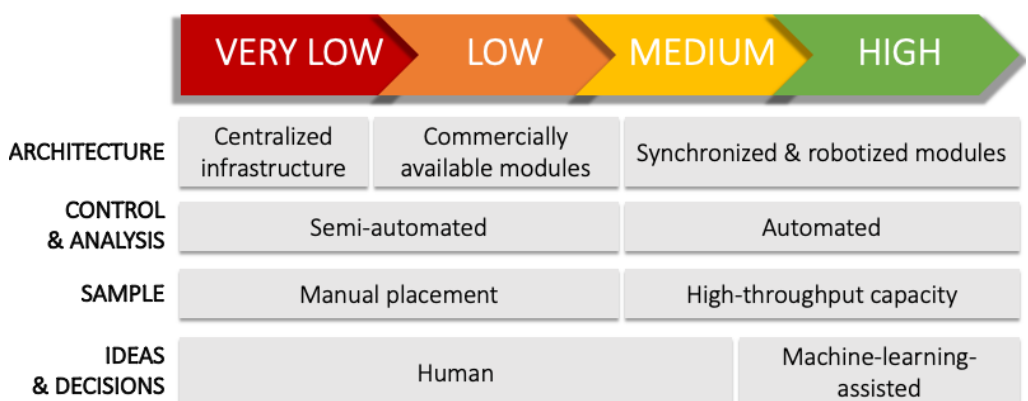

**Scheme 3**. Automation levels color-coded and classified according to different practical criteria for a MAP equipment.

Following the automation criteria above, some techniques necessary to measure key properties in photo(thermal)catalysis have been approximately labeled in Scheme 4 . On the 'very low' automation range, there are techniques based on X-ray, neutron and electron beams which are typically costly and are confined to centralized infrastructures. In parallel, laboratory equipment commonly used for solid state materials synthesis protocols usually have poor digital interfaces that challenge their integration into a computer-controlled setup.

On the 'low' automation range, we have included techniques such as spectroscopy or liquid and solid handling because several commercial modules are available. With these techniques, nevertheless, running automated measurement and analysis workflows of multiple samples is not always possible, as it is increasingly common in chromatography, electrochemistry or photochemistry, labeled as 'medium' automated. In the same 'medium' category, we have included chemical and physical vapor deposition, and ink-jet printing techniques, which offer the possibility to prepare combinatorial compositional libraries of different materials. In terms of reactor design, many aspects relying on customized parts and fine motor control can be difficult to automate. However, for testing materials used to build the reactors and specific reactor configurations, automation appears possible with computational modeling and 3D printing.[16] Therefore, in the short term, MAPs for gas-phase heterogeneous $CO_2$ photo(thermal)catalysis could be developed from 'low' and 'medium' automated techniques, while machine learning could be used to make the most out of the data from 'very low' automated techniques combined with theoretical data and available datasets.

| Descriptor type | Technique | Level of automation as defined in Scheme 3 |
|---|---|---|
| Materials Synthesis | Spin and dip coating | LOW |
| | Chemical vapor deposition (CVD) | MEDIUM |
| | Physical Vapor Deposition (PVD) | MEDIUM |
| | Air and inert atmosphere liquid and solid handling | LOW |
| | Thermal, Electrochemical, Photochemical | VERY LOW |
| Reactor Design | CAD and 3D printing | MEDIUM |
| Crystal Structure, Phase Analysis | Powder and single crystal X-ray diffraction (PXRD-SCXRD) | LOW |
| Morphology | Transmission electron microscopy (TEM) | VERY LOW |
| | Scanning electron microscopy – Energy dispersive X-ray (SEM-EDX) | VERY LOW |
| Elemental Composition | Elemental analysis, X-ray absorption and emission spectroscopies (XES, XAS) | VERY LOW |
| | X-ray photoelectron spectroscopy (XPS) | VERY LOW |
| Electrical, Optical, Magnetic, Vibrational | Spectroscopy (Impedance, UV-Vis, UPS, Raman, Infrared, ESR, SQUID) | LOW |
| | Electrochemistry | MEDIUM |
| Surface Area, Pore Size, Pore Volume | Brunauer-Emmett-Teller (BET) | LOW |
| Reaction Rate, Selectivity, Conversion, Surface Chemistry, Mechanism | Time-resolved spectroscopies (UV-Vis, Raman, IR) | LOW |
| Chemical, Thermal, Photochemical Stability | Mass spectrometry (MS), thermogravimetric analysis (TGA) | LOW |
| Energy Efficiency | Chromatography | MEDIUM |

**Scheme 4** . Typical level of automation of key experimental techniques considered necessary for a gas-phase heterogeneous $CO_2$ photo(thermal)catalyst preparation and characterization MAP, color-coded criteria as in Scheme 3 .

On the way towards building efficient machine-learning-assisted robotized multi-modular MAPs, we identify two trends in the materials chemistry, science, and engineering community: (1) developing high-throughput automation approaches and (2) proving the applicability of machine learning to interpret experimental data. In the case of high-throughput automation approaches, there are a few examples of characterized combinatorial depositions of photo- and electro-catalyst materials, which are mostly based on one experimental preparation-characterization-interpretation loop sometimes followed by a second loop testing the best candidates. For example, the photo(electro)catalytic performance of ink-jet-printed semiconductor metal oxides towards water splitting was evaluated by Parkinson et al.[22–24] and Lewis et al.,[25] while Parkin et al.[26] focused on metal oxides prepared by chemical vapor deposition. Both Parkinson and Parkin measured the optical response of a chemical probe at the photocatalyst surface, while Lewis used a photocurrent measurement setup with a translation stage. On the other hand, Gregoire et al.[27] and Ludwig et al.[28] used a scanning droplet cell to execute high-throughput electrochemical measurements of electrocatalysts. Gregoire then moved further to build a high-throughput synthesis and characterization setup for photoactive materials based on X-ray fluorescence (XRF) and diffraction (XRD), ultraviolet-visible spectroscopy and photoelectrochemical characterization.[29–31] This research could be enhanced by further synchronizing and automating different measurements and their corresponding analysis, and by applying machine learning. The latter has been applied to material synthesis prediction based on the available literature[32–34] and theoretical databases.[35–38] For reactor design, there have been a number of recent examples where photoreactors were constructed via 3D[39,40] and 4D[41] printing, and integration of such techniques a MAP would likely be highly beneficial. In parallel, machine learning has also been used to interpret the results of complex characterization data, mostly related to 'very low' automated techniques.[42–46] Table 3 summarizes relevant examples of machine learning methods applied to interpret characterization data from several key techniques featured in Scheme 4 which shall be used together with popular databases.[47–63] Integrating these methods into MAPs could facilitate the extraction and interpretation of experimental data.

Robotic automation allows the non-stop performance of laborious and repetitive procedural steps within experiments (e.g. tasks that would otherwise tire or not be possible with a human researcher), and therefore forms a critical component of MAPs. In the spirit of accelerated discovery, we would like to briefly add that, since physically moving the relevant components through space takes time (and may be time consuming in certain cases), perhaps it would be beneficial to consider innovative ways to reduce physical movements and the number of moving parts in a MAP (e.g. such as miniaturizing the experimental setup) to further contribute to the acceleration of experimentation steps and ultimately the rate of discovery.

| Technique | Examples of machine-learning-assisted analysis |
|---|---|
| X-ray diffraction (XRD) | ● Classification of crystal structure using a convolutional neural network, 2017, *Park et al.*[64]<br>● Symmetry prediction and knowledge discovery from X-ray diffraction patterns using an interpretable machine learning approach, 2020, *Suzuki et al.*[65]<br>● Emergence and distinction of classes in XRD data via machine learning, 2019, *Royse et al.*[66]<br>● Automating crystal-structure phase mapping by combining deep learning with constraint reasoning, *Gomes et al.*[67,68] |
| Transmission electron microscopy (TEM) | ● Deep learning of atomically resolved scanning transmission electron microscopy images: chemical identification and tracking local transformations, 2017, *Ziatdinov et al.*[43]<br>● Deep learning detection of nanoparticles and multiple object tracking of their dynamic evolution during in situ ETEM studies, 2022, *Faraz et al.*[69]<br>● Statistical characterization of the morphologies of nanoparticles through machine learning based electron microscopy image analysis, 2020, *Lee et al.*[70]<br>● Machine learning in scanning transmission electron microscopy, 2022, *Kalinin et al.* (Review)[45] |
| Scanning electron microscopy – Energy dispersive X-ray (SEM-EDX) | ● Neural network for nanoscience scanning electron microscope image recognition, 2017, *Modarres et al.*[71]<br>● Retrieving the quantitative chemical information at nanoscale from scanning electron microscope energy dispersive X-ray measurements by machine learning, 2017, *Jany et al.*[72]<br>● Application of machine learning techniques in mineral classification for scanning electron microscopy - energy dispersive X-ray spectroscopy (SEM-EDS), 2021, *Li et al.*[73] |
| X-ray emission and Absorption Spectroscopies (XES, XAS, etc.) | ● Extraction of physical parameters from X-ray spectromicroscopy data using machine learning, 2018, *Suzuki et al.*[74]<br>● Machine-Learning X-Ray absorption spectra to quantitative accuracy, 2020, *Carbone et al.*[75]<br>● "Inverting" X-ray absorption spectra of catalysts by machine learning in search for activity descriptors, 2019, *Timoshenko et al.* (Review)[46] |
| Spectroscopy (UV-Vis) | ● Machine learning of optical properties of materials – predicting spectra from images and images from Spectra, 2018, *Stein et al.*[76]<br>● Multi-component background learning automates signal detection for spectroscopic data, 2019, *Ament et al.*[77]<br>● Machine learning enhanced spectroscopic analysis: towards autonomous chemical mixture characterization for rapid process optimization, 2021, *Angulo et al.*[78] |
| Brunauer-Emmett-Teller (BET) | ● Beyond the BET analysis: the surface area prediction of nanoporous materials using a machine learning method, 2020, *Datar et al.*[79] |

**Table 1** . Examples of machine-learning-assisted analysis categorized by technique.

**Using artificial intelligence to accelerate catalyst discovery and reactor design**

How might the artificial intelligence component of a MAP for gas-phase heterogeneous $CO_2$ photo(thermal)catalysis look like? The specific problem lies in optimizing catalyst efficiency, selectivity, and stability, as discussed earlier. However, the size of the chemical and descriptor space limits the effectiveness of random trials and high-throughput experimentation for materials discovery, which is a process that ranges from exploring syntheses to optimizing properties. In the case of the chemical space, Walsh and coworkers estimated that the total number of two, three and four component materials that can potentially be synthesized from the first 103 elements in the periodic table are $1.5 \cdot 10^4$, $3.2 \cdot 10^7$, and $3.2 \cdot 10^{10}$ respectively, even after imposing electronegativity and charge balancing rules.[80,81] In contrast, the materials science community has reported only about $2 \cdot 10^5$ inorganic materials in the past century alone, a tiny fraction of this estimate of the chemical space.[82] Within these estimates, there would be millions of additional modificiations possible via non-stoichiometry. Practically, as a starting point it would be desirable to limit the search space to earth-abundant (and therefore cheaper), non-toxic, and chemically stable elements, ideally resulting in a search space that is more managable (i.e. $10^3 - 10^4$ materials) and better defined. Additionally, since for each composition there could be hundreds of modifications via site substitutions, defect engineering, and non-stoichiometry, when designing a MAP one should also consider whether to address these additional modifications in the primary screening phase or include these extra dimensions in a subsequent search and optimization step.

To truly accelerate the pace of discovery and minimize the associated costs, MAPs would employ the information from the results they generate, just like how humans can learn and make decisions after evaluating the outcomes from prior ones. In addition, "discovery" can also be taken to mean not just finding new materials, but also identifying new applications and improvements on existing ones. This is where artificial intelligence comes into play in a MAP.[81,83] Statistical or machine learning algorithms, in particular, can be used to identify complex non-linear patterns in previous data and to make predictions on how to achieve desired outcomes.[84–87] Therefore, aside from having automated and high-throughput measurements and analyses to generate large and representative datasets, it is critical that a MAP have a ML-based control centre that can effectively process the generated results and guide the next set of experiments effectively based on the available data.

Heterogeneous photo(thermal)catalysis is a problem with a multidimensional input and output space. Ideally, all the relevant parameters in the input and output space should be considered in a search, but often that may not be practically feasible. If the dimensions of the input and output space in the ML process are to be restricted, any unreasonable assumptions for a working heterogeneous photocatalytic process should be avoided, using for example unsupervised machine learning or dimensionality reduction techniques to identify the most relevant descriptors. However, one should aim to provide as many relevant input descriptors as reasonably possible to best describe the problem and/or target properties at hand. Supervised ML methods based on neural networks, random forests or Bayesian optimization could then be particularly suited to model high-dimensionality problems including categorical and quantitative parameters, and identify optimal reaction conditions.[86–90] In any case, when designing an appropriate MAP, certain best practices to ensure the reliability and reproducibility of the results should continue to be considered.[91]

The application of machine learning for $CO_2$ conversion so far has primarily focused on catalyst materials for electrocatalysis,[92–99] while machine learning as applied towards chemical reactor design has been studied widely on traditional chemical reactors.[100–104] For photochemical reactors, a recent study investigated how the structure of a compound parabolic collector and the operating conditions of the photoreactor can together affect hydrogen production.[105] By using optimization algorithms, the authors were able to increase the energy conversion efficiencies by more than three fold. In addition, high-throughput computational screening coupled with machine learning could be an important component of a MAP by pre-screening materials and conditions, or by generating complementary theoretical descriptors.[81,83] In $CO_2$ reduction, this approach has been applied to electrocatalysts using algorithms trained on first-principles simulations such as intermediate adsorption energy changes, which are correlated to the catalyst activity.[82,96,106–112] Integrating the *in-silico* and experimental discovery and optimization processes for both catalyst and reactor design will likely be crucial for accelerated development of $CO_2$ heterogeneous photo(thermal)catalysis.

A potential MAP would integrate both machine learning and automated experimental techniques into one synergistic workflow. In Scheme 5, we briefly outline a MAP concept for photo(thermal)catalyst discovery. For photoreactors, the materials and components of experimentation and testing would differ, but the main pillars and the overall workflow remains the same. While the focus here is on $CO_2$ heterogeneous photo(thermal)catalysis, we note that the MAP concept outlined below can be generalized beyond $CO_2$ heterogeneous photo(thermal)catalysis for other gases and applications for constructing a waste-to-resource framework.

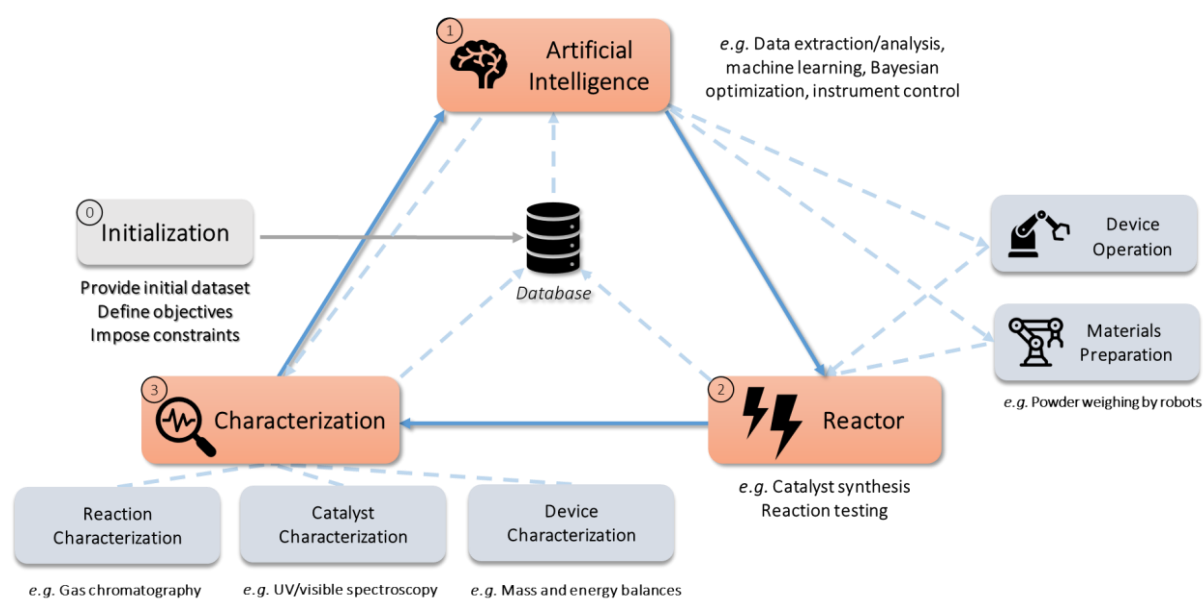

**Scheme 5.** A proposed MAP for photo(thermal)catalyst discovery. Solid blue arrows indicate the direction of workflow between the main pillars that form the operation cycle of a MAP (orange boxes), and dashed blue arrows indicate the parts that receive instructions from the AI-driven system on how to operate.

**Step 0 - Initialization.** For the cycle to begin, there needs to be a starting set of data from which the AI algorithm does its first round of learning and optimizing. This can be obtained as new data specifically for such initialization or as existing data, featurized in a way relevant to the problem at hand, from an available dataset. In any case, no "big data" is needed and 70 - 100 data points would likely be sufficient (e.g., the ARES system[113] for growing carbon nanotubes used only 84). In addition to the initial dataset, one should also define the objectives for the search (e.g. convergence criteria) and impose any relevant constraints on the search space (e.g., ranges for temperature, pressure, etc.).

**Step 1 - Artificial Intelligence.** This is the "AI-based control centre" (e.g., the brain) of the entire system. It extracts and analyzes data, learns from it via specific ML algorithms (e.g., data from a previous round would serve as feedback), crunches the numbers to estimate from available information where the best outcome for the next round might be, and plans the next set of experiments based on that information. Incorporating active learning, which can query a human on labeling how good the data is, could also be extremely beneficial here, since it not only allows the learning process to adapt to new data on the fly, but also reduces the number of samples needed to complete a specific learning task.

In terms of the ML models, interpretability is critical, and this may be one of the major challenges in developing a MAP for photo(thermal)catalysis. Typical ML algorithms are data-driven, which means it simply applies complex statistics to fit observational data. In such cases, expert domain knowledge in featuring the raw input (i.e. compositional, structural, morphological) data for training and selecting the target quantity would be essential to produce reliable and interpretable models. With a lot of physics and chemistry yet to be uncovered for surface localized photocatalysis and photothermal catalysis, advances on basic research in the area, the insights obtained on the relevant physiochemical trends of such processes from the results of data-driven models, and perhaps even newly developed models geared towards scientific understanding in the future,[114] can be used to help create physics (and/or chemistry) informed ML models (e.g. ML models where constraints from established physicochemical laws are imposed on an ML algorithm to improve its performance and interpretability)[115] for application in a MAP. Constructing interpretable ML models and predictions within physics and chemistry is not new,[65,67,72,74] but for photo(thermal)catalysis and its associated opto-electronic/opto-chemical engineering, to our knowledge it is essentially non-existent, and thus represents an endeavor ripe for exploration.

**Step 2 - Experimentation**. Interfaced with the relevant robotic and computer components, this is where all the relevant ingredients are brought together for testing and where the procedures for testing are carried out. For photocatalyst discovery, this can involve reactors for catalyst synthesis and reaction testing. This step receives instructions from the AI control centre and turns them into physical action on preparing catalysts and reagents, running reactions, and controlling experimental conditions. The components involved would be highly customized to the problem at hand and, for more specialized synthesis techniques, miniaturizing and/or interfacing the appropriate machinery for automation and integration into a MAP may be required. The experimental procedure (e.g., the instructions received from the AI control centre ) should be standardized, consistent, and generalizable so that it can be reproduced elsewhere when necessary. Finally, the outcomes of this step, in analog and physical form, is passed on to be read by components of the next pillar where it is digitized and quantified.

**Step 3 - Characterization**. Through the relevant characterization techniques, outcomes of the previous step are monitored and quantified. This could focus on the reaction being catalyzed (e.g., product quantity and distribution) and/or the catalyst materials (e.g., optical absorption properties, degradation/stability, catalyst intermediates). In addition, the efficiency of the device(s) used in the previous step can also be measured and quantified. The relevant instruments here would also receive instructions from AI on how to operate based on the reaction conditions stipulated in the previous step. Finally, the information produced in this step would be fed into the central database for the AI to process, analyze, and make decisions on for the next learning cycle.

All together, following initialization, the MAP would cycle through Steps 1 - 3 until it reaches a certain convergence threshold for one or more specified properties. What a MAP would do is akin to humans learning from initial data and trying to improve on it until some criteria is reached, except for the decision making in a MAP being guided by statistical models and automated through computer programs and robotics.

When combined with high-throughput experimentation and computation, a MAP with a well-defined objective, along with an interpretable and reliable ML model would not only be able to offer new discoveries in a shortened timeframe, but also be able to provide new insight on the relevant physio-chemical theories that are currently unknown or difficult to quantify.[116] More importantly, however, we should always be cautious and critically evaluate the process and outcomes of employing the relevant ML models, treating it not as a black box but conscientiously as a tool for accelerating meaningful discoveries.

**Up-Scaling Beyond the Lab**

Developing a commercial catalytic process requires many levels of development, many of which can be interdependent and interconnected. An efficient catalyst identified at the materials discovery phase may often become inapplicable beyond the domains of the research lab it was prepared and tested in, and a "research catalyst" is often very different from a "technical catalyst"[4] operating commercially. The latter, for example, often requires a variety of additives and formats, and health and safety protocols, for operating in industrial scale reactors. The problem of bringing a catalyst from lab-to-market is a science in itself[4] and is significant enough to own its own field. It is also a domain that we believe could benefit from the assistance of AI/ML.

Societal problems often cross many boundaries and are therefore highly interdisciplinary. When working in a research lab, it can be easy to fall into the trap of assuming that certain applications can be extrapolated or inferred from the results and conclusions of a particular study. For making predictions with ML models, it is commonly known as good practice to only interpolate and not extrapolate because it can be extremely risky and often nonsensical to make predictions beyond the domains from which the statistical model is trained on. In terms of developing materials for a certain application with MAPs, perhaps it might be useful to try and follow the concept of interpolation in ML models. In other words, consider all relevant components of a problem across the appropriate scales together in parallel, instead of addressing individual fragments of a larger problem separately and then trying to put them together after. The traditional way of splitting a problem into fragments and addressing each individually is still useful in many cases, but for complex multidimensional problems, it may also present a barrier to offering a valuable solution when the whole problem is clearly much greater than the sum of its interconnected components. With the advent of powerful statistical models and modern computational infrastructure, it is becoming increasingly possible treat a complex problem as a whole.

In terms of heterogeneous $CO_2$ photo(thermal)catalysis, the approach could involve building a "model" trained on all the necessary descriptors/features that go into commercialization of a certain product, in parallel across all development steps from chemistry to engineering to economics and possibly beyond, and then use it to interpolate the best combination of features that would satisfy predefined targets for commercialization. In terms of the features and targets for training a ML model on, this would extend beyond photocatalyst and photoreactor design and performance to include ones such as cost of raw materials, additives required, $CO_2$ footprint of the catalytic process, by-products produced, recycling, and potential profit margin.

This is easier said than done and represents an enormous task, and at the same time would likely also involve more than one MAP working together in parallel. However, we believe it is still useful to put forward this concept here as a potential route to consider for solving a multilayer and multidimensional problem like heterogeneous $CO_2$ photo(thermal)catalysis. By collaborating across experts and stakeholders from academia, industry, and government, perhaps it may be possible to integrate all the necessary information into one place for accelerating the development of applications to address critical problems in society.

**Conclusion**

Heterogeneous $CO_2$ photo(thermal)catalysis is an exciting challenge, one with the potential to transform the traditional thermal-based chemical technologies that currently dominate the catalytic industry. With increasing demand for sustainable fuels and feedstocks to help curb climate change, we look forward to seeing how MAPs can be designed and applied towards this underexplored domain to help address the ever-increasing societal problem with one of the most abundant natural resources around - solar energy.

**Author contributions**

A.W. provided the abstract, introduction, the last paragraph of the 2nd section, part of the 3rd section, the 4th section, the conclusion, and the organization/editing of the final document for submission. C.B.-G. provided the remainder of the 2nd section and was involved in feedback and editing the initial portions written by A.W. and S.G.H.K., who contributed to the 3rd section. All the authors mentioned so far were involved in preparing Scheme 5. G.A.O conceived the idea for writing this manuscript, and along with A.A.-G. supervised the project and provided feedback, edits, and directions.

**Declaration of Interests**

The authors have no conflicts of interest to declare.

**Acknowledgements**

G.A.O. and A.W. acknowledge funding from the Natural Sciences and Engineering Council of Canada (NSERC). C.B.-G. acknowledges funding from a M.S.C.A. Postdoctoral Fellowship grant (101064374). A.A.-G. acknowledges support from the Canada Industrial Research Chairs Program, and the Canada 150 Research Chairs Program, as well as funding from the National Research Council of Canada.

**References**

(1) Dong, Y.; Duchesne, P.; Mohan, A.; Ghuman, K. K.; Kant, P.; Hurtado, L.; Ulmer, U.; Loh, J. Y. Y.; Tountas, A. A.; Wang, L.; Jelle, A.; Xia, M.; Dittmeyer, R.; Ozin, G. A. Shining Light on $CO_2$: From Materials Discovery to Photocatalyst, Photoreactor and Process Engineering. *Chem. Soc. Rev.* **2020**, *49* (16), 5648–5663. https://doi.org/10.1039/D0CS00597E.
(2) Docao, S.; Koirala, A. R.; Kim, M. G.; Hwang, I. C.; Song, M. K.; Yoon, K. B. Solar Photochemical–Thermal Water Splitting at 140 °C with Cu-Loaded $TiO_2$. *Energy Environ. Sci.* **2017**, *10* (2), 628–640. https://doi.org/10.1039/C6EE02974D.
(3) Wang, A.; Tang, S.; Lee, M.; Ozin, G.A. Unpublished Work, 2022.
(4) Mitchell, S.; Michels, N.-L.; Pérez-Ramírez, J. From Powder to Technical Body: The Undervalued Science of Catalyst Scale Up. *Chem. Soc. Rev.* **2013**, *42* (14), 6094. https://doi.org/10.1039/c3cs60076a.
(5) Seifrid, M.; Hattrick-Simpers, J.; Aspuru-Guzik, A.; Kalil, T.; Cranford, S. Reaching Critical MASS: Crowdsourcing Designs for the next Generation of Materials Acceleration Platforms. *Matter* **2022**, *5* (7), 1972–1976. https://doi.org/10.1016/j.matt.2022.05.035.
(6) Chueh, W. C.; Falter, C.; Abbott, M.; Scipio, D.; Furler, P.; Haile, S. M.; Steinfeld, A. High-Flux Solar-Driven Thermochemical Dissociation of CO2 and H2O Using Nonstoichiometric Ceria. *Science* **2010**, *330* (6012), 1797–1801. https://doi.org/10.1126/science.1197834.
(7) Scheffe, J. R.; Steinfeld, A. Oxygen Exchange Materials for Solar Thermochemical Splitting of H2O and CO2: A Review. *Materials Today* **2014**, *17* (7), 341–348. https://doi.org/10.1016/j.mattod.2014.04.025.
(8) *Carbon-neutral fuel made from sunlight and air*. https://ethz.ch/en/news-and-events/eth-news/news/2019/06/pr-solar-mini-refinery.html.
(9) Lufthansa Group. *Flying with sunlight*. https://www.lufthansagroup.com/en/newsroom/releases/flying-with-sunlight.html.
(10) Zhang, Z.; Mao, C.; Meira, D. M.; Duchesne, P. N.; Tountas, A. A.; Li, Z.; Qiu, C.; Tang, S.; Song, R.; Ding, X.; Sun, J.; Yu, J.; Howe, J. Y.; Tu, W.; Wang, L.; Ozin, G. A. New Black Indium Oxide—Tandem Photothermal CO2-H2 Methanol Selective Catalyst. *Nat Commun* **2022**, *13* (1), 1512. https://doi.org/10.1038/s41467-022-29222-7.
(11) Wang, L.; Dong, Y.; Yan, T.; Hu, Z.; Jelle, A. A.; Meira, D. M.; Duchesne, P. N.; Loh, J. Y. Y.; Qiu, C.; Storey, E. E.; Xu, Y.; Sun, W.; Ghoussoub, M.; Kherani, N. P.; Helmy, A. S.; Ozin, G. A. Black Indium Oxide a Photothermal CO2 Hydrogenation Catalyst. *Nat Commun* **2020**, *11* (1), 2432. https://doi.org/10.1038/s41467-020-16336-z.

(12) Ghuman, K. K.; Hoch, L. B.; Szymanski, P.; Loh, J. Y. Y.; Kherani, N. P.; El-Sayed, M. A.; Ozin, G. A.; Singh, C. V. Photoexcited Surface Frustrated Lewis Pairs for Heterogeneous Photocatalytic $CO_2$ Reduction. *J. Am. Chem. Soc.* **2016**, *138* (4), 1206–1214. https://doi.org/10.1021/jacs.5b10179.

(13) Hurtado, L.; Mohan, A.; Ulmer, U.; Natividad, R.; Tountas, A. A.; Sun, W.; Wang, L.; Kim, B.; Sain, M. M.; Ozin, G. A. Solar $CO_2$ Hydrogenation by Photocatalytic Foams. *Chemical Engineering Journal* **2022**, *435*, 134864. https://doi.org/10.1016/j.cej.2022.134864.

(14) Wang, B.; Wang, X.; Lu, L.; Zhou, C.; Xin, Z.; Wang, J.; Ke, X.; Sheng, G.; Yan, S.; Zou, Z. Oxygen-Vacancy-Activated $CO_2$ Splitting over Amorphous Oxide Semiconductor Photocatalyst. *ACS Catal.* **2018**, *8* (1), 516–525. https://doi.org/10.1021/acscatal.7b02952.

(15) Ghoussoub, M.; Xia, M.; Duchesne, P. N.; Segal, D.; Ozin, G. Principles of Photothermal Gas-Phase Heterogeneous $CO_2$ Catalysis. *Energy Environ. Sci.* **2019**, *12* (4), 1122–1142. https://doi.org/10.1039/C8EE02790K.

(16) Kant, P.; Trinkies, L. L.; Gensior, N.; Fischer, D.; Rubin, M.; Ozin, G. A.; Dittmeyer, R. Isophotonic Reactor for the Precise Determination of Quantum Yields in Gas, Liquid, and Multi-Phase Photoreactions. *Chemical Engineering Journal* **2023**, *452*, 139204. https://doi.org/10.1016/j.cej.2022.139204.

(17) Schreiber, A.; Peschel, A.; Hentschel, B.; Zapp, P. Life Cycle Assessment of Power-to-Syngas: Comparing High Temperature Co-Electrolysis and Steam Methane Reforming. *Front. Energy Res.* **2020**, *8*, 533850. https://doi.org/10.3389/fenrg.2020.533850.

(18) Gong, E.; Ali, S.; Hiragond, C. B.; Kim, H. S.; Powar, N. S.; Kim, D.; Kim, H.; In, S.-I. Solar Fuels: Research and Development Strategies to Accelerate Photocatalytic $CO_2$ Conversion into Hydrocarbon Fuels. *Energy Environ. Sci.* **2022**, *15* (3), 880–937. https://doi.org/10.1039/D1EE02714J.

(19) Ozin, G. Accelerated Optochemical Engineering Solutions to $CO_2$ Photocatalysis for a Sustainable Future. *Matter* **2022**, *5* (9), 2594–2614. https://doi.org/10.1016/j.matt.2022.07.033.

(20) Häse, F.; Roch, L. M.; Aspuru-Guzik, A. Next-Generation Experimentation with Self-Driving Laboratories. *Trends in Chemistry* **2019**, *1* (3), 282–291. https://doi.org/10.1016/j.trechm.2019.02.007.

(21) Flores-Leonar, M. M.; Mejía-Mendoza, L. M.; Aguilar-Granda, A.; Sanchez-Lengeling, B.; Tribukait, H.; Amador-Bedolla, C.; Aspuru-Guzik, A. Materials Acceleration Platforms: On the Way to Autonomous Experimentation. *Current Opinion in Green and Sustainable Chemistry* **2020**, *25*, 100370. https://doi.org/10.1016/j.cogsc.2020.100370.

(22) Woodhouse, M.; Parkinson, B. A. Combinatorial Discovery and Optimization of a Complex Oxide with Water Photoelectrolysis Activity. *Chem. Mater.* **2008**, *20* (7), 2495–2502. https://doi.org/10.1021/cm703099j.

(23) Woodhouse, M.; Herman, G. S.; Parkinson, B. A. Combinatorial Approach to Identification of Catalysts for the Photoelectrolysis of Water. *Chem. Mater.* **2005**, *17* (17), 4318–4324. https://doi.org/10.1021/cm050546q.

(24) Seley, D.; Ayers, K.; Parkinson, B. A. Combinatorial Search for Improved Metal Oxide Oxygen Evolution Electrocatalysts in Acidic Electrolytes. *ACS Comb. Sci.* **2013**, *15* (2), 82–89. https://doi.org/10.1021/co300086h.

(25) Katz, J. E.; Gingrich, T. R.; Santori, E. A.; Lewis, N. S. Combinatorial Synthesis and High-Throughput Photopotential and Photocurrent Screening of Mixed-Metal Oxides for Photoelectrochemical Water Splitting. *Energy Environ. Sci.* **2009**, *2* (1), 103–112. https://doi.org/10.1039/B812177J.

(26) Kafizas, A.; Parkin, I. P. Combinatorial Atmospheric Pressure Chemical Vapor Deposition (CAPCVD): A Route to Functional Property Optimization. *J. Am. Chem. Soc.* **2011**, *133* (50), 20458–20467. https://doi.org/10.1021/ja208633g.

(27) Gregoire, J. M.; Xiang, C.; Liu, X.; Marcin, M.; Jin, J. Scanning Droplet Cell for High Throughput Electrochemical and Photoelectrochemical Measurements. *Review of Scientific Instruments* **2013**, *84* (2), 024102. https://doi.org/10.1063/1.4790419.

(28) Batchelor, T. A. A.; Löffler, T.; Xiao, B.; Krysiak, O. A.; Strotkötter, V.; Pedersen, J. K.; Clausen, C. M.; Savan, A.; Li, Y.; Schuhmann, W.; Rossmeisl, J.; Ludwig, A. Complex-Solid-Solution Electrocatalyst Discovery by Computational Prediction and High-Throughput Experimentation**. *Angew. Chem. Int. Ed.* **2021**, *60* (13), 6932–6937. https://doi.org/10.1002/anie.202014374.

(29) Gregoire, J. M.; Xiang, C.; Mitrovic, S.; Liu, X.; Marcin, M.; Cornell, E. W.; Fan, J.; Jin, J. Combined Catalysis and Optical Screening for High Throughput Discovery of Solar Fuels Catalysts. *J. Electrochem. Soc.* **2013**, *160* (4), F337–F342. https://doi.org/10.1149/2.035304jes.

(30) Yan, Q.; Yu, J.; Suram, S. K.; Zhou, L.; Shinde, A.; Newhouse, P. F.; Chen, W.; Li, G.; Persson, K. A.; Gregoire, J. M.; Neaton, J. B. Solar Fuels Photoanode Materials Discovery by Integrating High-Throughput Theory and Experiment. *Proc. Natl. Acad. Sci. U.S.A.* **2017**, *114* (12), 3040–3043. https://doi.org/10.1073/pnas.1619940114.

(31) Noh, J.; Kim, S.; Gu, G. ho; Shinde, A.; Zhou, L.; Gregoire, J. M.; Jung, Y. Unveiling New Stable Manganese Based Photoanode Materials *via* Theoretical High-Throughput Screening and Experiments. *Chem. Commun.* **2019**, *55* (89), 13418–13421. https://doi.org/10.1039/C9CC06736A.

(32) Jensen, Z.; Kim, E.; Kwon, S.; Gani, T. Z. H.; Román-Leshkov, Y.; Moliner, M.; Corma, A.; Olivetti, E. A Machine Learning Approach to Zeolite Synthesis Enabled by Automatic Literature Data Extraction. *ACS Cent. Sci.* **2019**, *5* (5), 892–899. https://doi.org/10.1021/acscentsci.9b00193.

(33) Kim, E.; Huang, K.; Saunders, A.; McCallum, A.; Ceder, G.; Olivetti, E. Materials Synthesis Insights from Scientific Literature via Text Extraction and Machine Learning. *Chem. Mater.* **2017**, *29* (21), 9436–9444. https://doi.org/10.1021/acs.chemmater.7b03500.

(34) Kim, E.; Huang, K.; Tomala, A.; Matthews, S.; Strubell, E.; Saunders, A.; McCallum, A.; Olivetti, E. Machine-Learned and Codified Synthesis Parameters of Oxide Materials. *Sci Data* **2017**, *4* (1), 170127. https://doi.org/10.1038/sdata.2017.127.

(35) Huo, H.; Rong, Z.; Kononova, O.; Sun, W.; Botari, T.; He, T.; Tshitoyan, V.; Ceder, G. Semi-Supervised Machine-Learning Classification of Materials Synthesis Procedures. *npj Comput Mater* **2019**, *5* (1), 62. https://doi.org/10.1038/s41524-019-0204-1.

(36) Shi, Y.-F.; Kang, P.-L.; Shang, C.; Liu, Z.-P. Methanol Synthesis from $CO_2$/CO Mixture on Cu–Zn Catalysts from Microkinetics-Guided Machine Learning Pathway Search. *J. Am. Chem. Soc.* **2022**, *144* (29), 13401–13414. https://doi.org/10.1021/jacs.2c06044.

(37) Takahashi, K.; Tanaka, Y. Material Synthesis and Design from First Principle Calculations and Machine Learning. *Computational Materials Science* **2016**, *112*, 364–367. https://doi.org/10.1016/j.commatsci.2015.11.013.

(38) Shimizu, R.; Kobayashi, S.; Watanabe, Y.; Ando, Y.; Hitosugi, T. Autonomous Materials Synthesis by Machine Learning and Robotics. *APL Materials* **2020**, *8* (11), 111110. https://doi.org/10.1063/5.0020370.

(39) Phang, S. J.; Wong, V.-L.; Cheah, K. H.; Tan, L.-L. 3D-Printed Photoreactor with Robust g-C3N4 Homojunction Based Thermoset Coating as a New and Sustainable Approach for Photocatalytic Wastewater Treatment. *Journal of Environmental Chemical Engineering* **2021**, *9* (6), 106437. https://doi.org/10.1016/j.jece.2021.106437.

(40) Schiel, F.; Peinsipp, C.; Kornigg, S.; Böse, D. A 3D-Printed Open Access Photoreactor Designed for Versatile Applications in Photoredox- and Photoelectrochemical Synthesis**. *ChemPhotoChem* **2021**, *5* (5), 431–437. https://doi.org/10.1002/cptc.202000291.

(41) Hansen, A.; Renner, M.; Griesbeck, A. G.; Büsgen, T. From 3D to 4D Printing: A Reactor for Photochemical Experiments Using Hybrid Polyurethane Acrylates for Vat-Based Polymerization and Surface Functionalization. *Chem. Commun.* **2020**, *56* (96), 15161–15164. https://doi.org/10.1039/D0CC06512A.

(42) Noack, M. M.; Zwart, P. H.; Ushizima, D. M.; Fukuto, M.; Yager, K. G.; Elbert, K. C.; Murray, C. B.; Stein, A.; Doerk, G. S.; Tsai, E. H. R.; Li, R.; Freychet, G.; Zhernenkov, M.; Holman, H.-Y. N.; Lee, S.; Chen, L.; Rotenberg, E.; Weber, T.; Goc, Y. L.; Boehm, M.; Steffens, P.; Mutti, P.; Sethian, J. A. Gaussian Processes for Autonomous Data Acquisition at Large-Scale Synchrotron and Neutron Facilities. *Nat Rev Phys* **2021**, *3* (10), 685–697. https://doi.org/10.1038/s42254-021-00345-y.

(43) Ziatdinov, M.; Dyck, O.; Maksov, A.; Li, X.; Sang, X.; Xiao, K.; Unocic, R. R.; Vasudevan, R.; Jesse, S.; Kalinin, S. V. Deep Learning of Atomically Resolved Scanning Transmission Electron Microscopy Images: Chemical Identification and Tracking Local Transformations. *ACS Nano* **2017**, *11* (12), 12742–12752. https://doi.org/10.1021/acsnano.7b07504.

(44) Chen, Z.; Andrejevic, N.; Drucker, N. C.; Nguyen, T.; Xian, R. P.; Smidt, T.; Wang, Y.; Ernstorfer, R.; Tennant, D. A.; Chan, M.; Li, M. Machine Learning on Neutron and X-Ray Scattering and Spectroscopies. *Chem. Phys. Rev.* **2021**, *2* (3), 031301. https://doi.org/10.1063/5.0049111.

(45) Kalinin, S. V.; Ophus, C.; Voyles, P. M.; Erni, R.; Kepaptsoglou, D.; Grillo, V.; Lupini, A. R.; Oxley, M. P.; Schwenker, E.; Chan, M. K. Y.; Etheridge, J.; Li, X.; Han, G. G. D.; Ziatdinov, M.; Shibata, N.; Pennycook, S. J. Machine Learning in Scanning Transmission Electron Microscopy. *Nat Rev Methods Primers* **2022**, *2* (1), 11. https://doi.org/10.1038/s43586-022-00095-w.
(46) Timoshenko, J.; Frenkel, A. I. "Inverting" X-Ray Absorption Spectra of Catalysts by Machine Learning in Search for Activity Descriptors. *ACS Catal.* **2019**, *9* (11), 10192–10211. https://doi.org/10.1021/acscatal.9b03599.
(47) *International Center for Diffraction Data*. icdd.com.
(48) *Crystallography Open Database*. crystallography.net.
(49) *Pauling File – Inorganic Materials Database*. paulingfile.com.
(50) *Materials Data Bank*. www.materialsdatabank.org.
(51) *Pearson´s Crystal Data*. crystalimpact.com.
(52) *NIST Databases*. nist.gov.
(53) *AFLOW*. aflowlib.org.
(54) *Materials Project*. materialsproject.org.
(55) *Open Quantum Materials Database*. oqmd.org.
(56) *NREL Materials Database*. materials.nrel.gov.
(57) *NOMAD*. nomad-lab.eu.
(58) *CALPHAD*. thermocalc.com.
(59) *MatWeb*. matweb.com.
(60) *Citrination*. citrination.com.
(61) *Springer Materials*. materials.springer.com.
(62) *Total Materia*. totalmateria.com.
(63) *ICSD FIZ Karlsruche*. icsd.products.fiz-karlsruhe.de.
(64) Park, W. B.; Chung, J.; Jung, J.; Sohn, K.; Singh, S. P.; Pyo, M.; Shin, N.; Sohn, K.-S. Classification of Crystal Structure Using a Convolutional Neural Network. *IUCrJ* **2017**, *4* (4), 486–494. https://doi.org/10.1107/S205225251700714X.
(65) Suzuki, Y.; Hino, H.; Hawai, T.; Saito, K.; Kotsugi, M.; Ono, K. Symmetry Prediction and Knowledge Discovery from X-Ray Diffraction Patterns Using an Interpretable Machine Learning Approach. *Sci Rep* **2020**, *10* (1), 21790. https://doi.org/10.1038/s41598-020-77474-4.
(66) Royse, C.; Wolter, S. D.; Greenberg, J. A. Emergence and Distinction of Classes in XRD Data via Machine Learning. In *Anomaly Detection and Imaging with X-Rays (ADIX) IV*; Ashok, A., Gehm, M. E., Greenberg, J. A., Eds.; SPIE: Baltimore, United States, 2019; p 12. https://doi.org/10.1117/12.2519500.
(67) Chen, D.; Bai, Y.; Ament, S.; Zhao, W.; Guevarra, D.; Zhou, L.; Selman, B.; van Dover, R. B.; Gregoire, J. M.; Gomes, C. P. Automating Crystal-Structure Phase Mapping by Combining Deep Learning with Constraint Reasoning. *Nat Mach Intell* **2021**, *3* (9), 812–822. https://doi.org/10.1038/s42256-021-00384-1.
(68) Gomes, C. P.; Bai, J.; Xue, Y.; Björck, J.; Rappazzo, B.; Ament, S.; Bernstein, R.; Kong, S.; Suram, S. K.; van Dover, R. B.; Gregoire, J. M. CRYSTAL: A Multi-Agent AI System for Automated Mapping of Materials' Crystal Structures. *MRS Communications* **2019**, *9* (2), 600–608. https://doi.org/10.1557/mrc.2019.50.
(69) Faraz, K.; Grenier, T.; Ducottet, C.; Epicier, T. Deep Learning Detection of Nanoparticles and Multiple Object Tracking of Their Dynamic Evolution during in Situ ETEM Studies. *Sci Rep* **2022**, *12* (1), 2484. https://doi.org/10.1038/s41598-022-06308-2.
(70) Lee, B.; Yoon, S.; Lee, J. W.; Kim, Y.; Chang, J.; Yun, J.; Ro, J. C.; Lee, J.-S.; Lee, J. H. Statistical Characterization of the Morphologies of Nanoparticles through Machine Learning Based Electron Microscopy Image Analysis. *ACS Nano* **2020**, *14* (12), 17125–17133. https://doi.org/10.1021/acsnano.0c06809.
(71) Modarres, M. H.; Aversa, R.; Cozzini, S.; Ciancio, R.; Leto, A.; Brandino, G. P. Neural Network for Nanoscience Scanning Electron Microscope Image Recognition. *Sci Rep* **2017**, *7* (1), 13282. https://doi.org/10.1038/s41598-017-13565-z.
(72) Jany, B. R.; Janas, A.; Krok, F. Retrieving the Quantitative Chemical Information at Nanoscale from Scanning Electron Microscope Energy Dispersive X-Ray Measurements by Machine Learning. *Nano Lett.* **2017**, *17* (11), 6520–6525. https://doi.org/10.1021/acs.nanolett.7b01789.
(73) Li, C.; Wang, D.; Kong, L. Application of Machine Learning Techniques in Mineral Classification for Scanning Electron Microscopy - Energy Dispersive X-Ray Spectroscopy (SEM-EDS) Images.

*Journal of Petroleum Science and Engineering* **2021**, *200*, 108178. https://doi.org/10.1016/j.petrol.2020.108178.

(74) Suzuki, Y.; Hino, H.; Ueno, T.; Takeichi, Y.; Kotsugi, M.; Ono, K. Extraction of Physical Parameters from X-Ray Spectromicroscopy Data Using Machine Learning. *Microsc Microanal* **2018**, *24* (S2), 478–479. https://doi.org/10.1017/S1431927618014629.

(75) Carbone, M. R.; Topsakal, M.; Lu, D.; Yoo, S. Machine-Learning X-Ray Absorption Spectra to Quantitative Accuracy. *Phys. Rev. Lett.* **2020**, *124* (15), 156401. https://doi.org/10.1103/PhysRevLett.124.156401.

(76) Stein, H. S.; Guevarra, D.; Newhouse, P. F.; Soedarmadji, E.; Gregoire, J. M. Machine Learning of Optical Properties of Materials – Predicting Spectra from Images and Images from Spectra. *Chem. Sci.* **2019**, *10* (1), 47–55. https://doi.org/10.1039/C8SC03077D.

(77) Ament, S. E.; Stein, H. S.; Guevarra, D.; Zhou, L.; Haber, J. A.; Boyd, D. A.; Umehara, M.; Gregoire, J. M.; Gomes, C. P. Multi-Component Background Learning Automates Signal Detection for Spectroscopic Data. *npj Comput Mater* **2019**, *5* (1), 77. https://doi.org/10.1038/s41524-019-0213-0.

(78) Angulo, A.; Yang, L.; Aydil, E. S.; Modestino, M. A. Machine Learning Enhanced Spectroscopic Analysis: Towards Autonomous Chemical Mixture Characterization for Rapid Process Optimization. *Digital Discovery* **2022**, *1* (1), 35–44. https://doi.org/10.1039/D1DD00027F.

(79) Datar, A.; Chung, Y. G.; Lin, L.-C. Beyond the BET Analysis: The Surface Area Prediction of Nanoporous Materials Using a Machine Learning Method. *J. Phys. Chem. Lett.* **2020**, *11* (14), 5412–5417. https://doi.org/10.1021/acs.jpclett.0c01518.

(80) Davies, D. W.; Butler, K. T.; Jackson, A. J.; Morris, A.; Frost, J. M.; Skelton, J. M.; Walsh, A. Computational Screening of All Stoichiometric Inorganic Materials. *Chem* **2016**, *1* (4), 617–627. https://doi.org/10.1016/j.chempr.2016.09.010.

(81) Butler, K. T.; Davies, D. W.; Cartwright, H.; Isayev, O.; Walsh, A. Machine Learning for Molecular and Materials Science. *Nature* **2018**, *559* (7715), 547–555. https://doi.org/10.1038/s41586-018-0337-2.

(82) Noh, J.; Kim, J.; Stein, H. S.; Sanchez-Lengeling, B.; Gregoire, J. M.; Aspuru-Guzik, A.; Jung, Y. Inverse Design of Solid-State Materials via a Continuous Representation. *Matter* **2019**, *1* (5), 1370–1384. https://doi.org/10.1016/j.matt.2019.08.017.

(83) Walsh, A. The Quest for New Functionality. *Nature Chem* **2015**, *7* (4), 274–275. https://doi.org/10.1038/nchem.2213.

(84) Häse, F.; Roch, L. M.; Kreisbeck, C.; Aspuru-Guzik, A. Phoenics: A Bayesian Optimizer for Chemistry. *ACS Cent. Sci.* **2018**, *4* (9), 1134–1145. https://doi.org/10.1021/acscentsci.8b00307.

(85) Häse, F.; Aldeghi, M.; Hickman, R. J.; Roch, L. M.; Aspuru-Guzik, A. Gryffin: An Algorithm for Bayesian Optimization of Categorical Variables Informed by Expert Knowledge. *Applied Physics Reviews* **2021**, *8* (3), 031406. https://doi.org/10.1063/5.0048164.

(86) Häse, F.; Aldeghi, M.; Hickman, R. J.; Roch, L. M.; Christensen, M.; Liles, E.; Hein, J. E.; Aspuru-Guzik, A. Olympus: A Benchmarking Framework for Noisy Optimization and Experiment Planning. *Mach. Learn.: Sci. Technol.* **2021**, *2* (3), 035021. https://doi.org/10.1088/2632-2153/abedc8.

(87) Hickman, R. J.; Häse, F.; Roch, L. M.; Aspuru-Guzik, A. Gemini: Dynamic Bias Correction for Autonomous Experimentation and Molecular Simulation. **2021**. https://doi.org/10.48550/ARXIV.2103.03391.

(88) Maddox, W. J.; Balandat, M.; Wilson, A. G.; Bakshy, E. Bayesian Optimization with High-Dimensional Outputs. **2021**. https://doi.org/10.48550/ARXIV.2106.12997.

(89) Xu, D.; Shi, Y.; Tsang, I. W.; Ong, Y.-S.; Gong, C.; Shen, X. Survey on Multi-Output Learning. *IEEE Trans. Neural Netw. Learning Syst.* **2019**, 1–21. https://doi.org/10.1109/TNNLS.2019.2945133.

(90) Segal, M.; Xiao, Y. Multivariate Random Forests. *WIREs Data Mining Knowl Discov* **2011**, *1* (1), 80–87. https://doi.org/10.1002/widm.12.

(91) Artrith, N.; Butler, K. T.; Coudert, F.-X.; Han, S.; Isayev, O.; Jain, A.; Walsh, A. Best Practices in Machine Learning for Chemistry. *Nat. Chem.* **2021**, *13* (6), 505–508. https://doi.org/10.1038/s41557-021-00716-z.

(92) Masood, H.; Toe, C. Y.; Teoh, W. Y.; Sethu, V.; Amal, R. Machine Learning for Accelerated Discovery of Solar Photocatalysts. *ACS Catal.* **2019**, *9* (12), 11774–11787. https://doi.org/10.1021/acscatal.9b02531.

(93) Mazheika, A.; Wang, Y.-G.; Valero, R.; Viñes, F.; Illas, F.; Ghiringhelli, L. M.; Levchenko, S. V.; Scheffler, M. Artificial-Intelligence-Driven Discovery of Catalyst Genes with Application to CO2 Activation on Semiconductor Oxides. *Nat Commun* **2022**, *13* (1), 419. https://doi.org/10.1038/s41467-022-28042-z.

(94) Zhang, N.; Yang, B.; Liu, K.; Li, H.; Chen, G.; Qiu, X.; Li, W.; Hu, J.; Fu, J.; Jiang, Y.; Liu, M.; Ye, J. Machine Learning in Screening High Performance Electrocatalysts for $CO_2$ Reduction. *Small Methods* **2021**, *5* (11), 2100987. https://doi.org/10.1002/smtd.202100987.

(95) Xiang, S.; Huang, P.; Li, J.; Liu, Y.; Marcella, N.; Routh, P. K.; Li, G.; Frenkel, A. I. Solving the Structure of "Single-Atom" Catalysts Using Machine Learning – Assisted XANES Analysis. *Phys. Chem. Chem. Phys.* **2022**, *24* (8), 5116–5124. https://doi.org/10.1039/D1CP05513E.

(96) Chen, A.; Zhang, X.; Chen, L.; Yao, S.; Zhou, Z. A Machine Learning Model on Simple Features for $CO_2$ Reduction Electrocatalysts. *J. Phys. Chem. C* **2020**, *124* (41), 22471–22478. https://doi.org/10.1021/acs.jpcc.0c05964.

(97) Zhong, M.; Tran, K.; Min, Y.; Wang, C.; Wang, Z.; Dinh, C.-T.; De Luna, P.; Yu, Z.; Rasouli, A. S.; Brodersen, P.; Sun, S.; Voznyy, O.; Tan, C.-S.; Askerka, M.; Che, F.; Liu, M.; Seifitokaldani, A.; Pang, Y.; Lo, S.-C.; Ip, A.; Ulissi, Z.; Sargent, E. H. Accelerated Discovery of CO2 Electrocatalysts Using Active Machine Learning. *Nature* **2020**, *581* (7807), 178–183. https://doi.org/10.1038/s41586-020-2242-8.

(98) Malek, A.; Wang, Q.; Baumann, S.; Guillon, O.; Eikerling, M.; Malek, K. A Data-Driven Framework for the Accelerated Discovery of CO2 Reduction Electrocatalysts. *Front. Energy Res.* **2021**, *9*, 609070. https://doi.org/10.3389/fenrg.2021.609070.

(99) Saadetnejad, D.; Oral, B.; Can, E.; Yıldırım, R. Machine Learning Analysis of Gas Phase Photocatalytic CO2 Reduction for Hydrogen Production. *International Journal of Hydrogen Energy* **2022**, *47* (45), 19655–19668. https://doi.org/10.1016/j.ijhydene.2022.02.030.

(100) Babanezhad, M.; Behroyan, I.; Nakhjiri, A. T.; Marjani, A.; Rezakazemi, M.; Shirazian, S. High-Performance Hybrid Modeling Chemical Reactors Using Differential Evolution Based Fuzzy Inference System. *Sci Rep* **2020**, *10* (1), 21304. https://doi.org/10.1038/s41598-020-78277-3.

(101) Vasickaninova, A.; Bakosova, M. Neural Network Predictive Control Of A Chemical Reactor. In *ECMS 2009 Proceedings edited by J. Otamendi, A. Bargiela, J. L. Montes, L. M. Doncel Pedrera*; ECMS, 2009; pp 563–569. https://doi.org/10.7148/2009-0563-0569.

(102) Vilim, R.; Ibarra, L. Artificial Intelligence/Machine Learning Technologies for Advanced Reactors: Workshop Summary Report, 2022. https://publications.anl.gov/anlpubs/2022/04/174557.pdf.

(103) Zhou, Z.; Li, X.; Zare, R. N. Optimizing Chemical Reactions with Deep Reinforcement Learning. *ACS Cent. Sci.* **2017**, *3* (12), 1337–1344. https://doi.org/10.1021/acscentsci.7b00492.

(104) Neo, R. *Designing a Chemical Reactor with Data Science*. https://towardsdatascience.com/designing-a-chemical-reactor-with-data-science-9e2c714d2475.

(105) Ren, T.; Ma, T.; Liu, S.; Li, X. Bi-Level Optimization for the Energy Conversion Efficiency Improvement in a Photocatalytic-Hydrogen-Production System. *Energy* **2022**, *253*, 124138. https://doi.org/10.1016/j.energy.2022.124138.

(106) Artrith, N.; Kolpak, A. M. Understanding the Composition and Activity of Electrocatalytic Nanoalloys in Aqueous Solvents: A Combination of DFT and Accurate Neural Network Potentials. *Nano Lett.* **2014**, *14* (5), 2670–2676. https://doi.org/10.1021/nl5005674.

(107) Ma, X.; Li, Z.; Achenie, L. E. K.; Xin, H. Machine-Learning-Augmented Chemisorption Model for $CO_2$ Electroreduction Catalyst Screening. *J. Phys. Chem. Lett.* **2015**, *6* (18), 3528–3533. https://doi.org/10.1021/acs.jpclett.5b01660.

(108) Tran, K.; Ulissi, Z. W. Active Learning across Intermetallics to Guide Discovery of Electrocatalysts for CO2 Reduction and H2 Evolution. *Nat Catal* **2018**, *1* (9), 696–703. https://doi.org/10.1038/s41929-018-0142-1.

(109) Ulissi, Z. W.; Tang, M. T.; Xiao, J.; Liu, X.; Torelli, D. A.; Karamad, M.; Cummins, K.; Hahn, C.; Lewis, N. S.; Jaramillo, T. F.; Chan, K.; Nørskov, J. K. Machine-Learning Methods Enable Exhaustive Searches for Active Bimetallic Facets and Reveal Active Site Motifs for $CO_2$ Reduction. *ACS Catal.* **2017**, *7* (10), 6600–6608. https://doi.org/10.1021/acscatal.7b01648.

(110) Sun, Y.; Yang, G.; Wen, C.; Zhang, L.; Sun, Z. Artificial Neural Networks with Response Surface Methodology for Optimization of Selective CO2 Hydrogenation Using K-Promoted Iron

Catalyst in a Microchannel Reactor. *Journal of CO2 Utilization* **2018**, *24*, 10–21. https://doi.org/10.1016/j.jcou.2017.11.013.

(111) Thornton, A. W.; Winkler, D. A.; Liu, M. S.; Haranczyk, M.; Kennedy, D. F. Towards Computational Design of Zeolite Catalysts for $CO_2$ Reduction. *RSC Adv.* **2015**, *5* (55), 44361–44370. https://doi.org/10.1039/C5RA06214D.

(112) Günay, M. E.; Türker, L.; Tapan, N. A. Decision Tree Analysis for Efficient CO2 Utilization in Electrochemical Systems. *Journal of CO2 Utilization* **2018**, *28*, 83–95. https://doi.org/10.1016/j.jcou.2018.09.011.

(113) Nikolaev, P.; Hooper, D.; Webber, F.; Rao, R.; Decker, K.; Krein, M.; Poleski, J.; Barto, R.; Maruyama, B. Autonomy in Materials Research: A Case Study in Carbon Nanotube Growth. *npj Comput Mater* **2016**, *2* (1), 16031. https://doi.org/10.1038/npjcompumats.2016.31.

(114) Krenn, M.; Pollice, R.; Guo, S. Y.; Aldeghi, M.; Cervera-Lierta, A.; Friederich, P.; dos Passos Gomes, G.; Häse, F.; Jinich, A.; Nigam, A.; Yao, Z.; Aspuru-Guzik, A. On Scientific Understanding with Artificial Intelligence. *Nat Rev Phys* **2022**, *4* (12), 761–769. https://doi.org/10.1038/s42254-022-00518-3.

(115) Karniadakis, G. E.; Kevrekidis, I. G.; Lu, L.; Perdikaris, P.; Wang, S.; Yang, L. Physics-Informed Machine Learning. *Nat Rev Phys* **2021**, *3* (6), 422–440. https://doi.org/10.1038/s42254-021-00314-5.

(116) Ren, F.; Ward, L.; Williams, T.; Laws, K. J.; Wolverton, C.; Hattrick-Simpers, J.; Mehta, A. Accelerated Discovery of Metallic Glasses through Iteration of Machine Learning and High-Throughput Experiments. *Sci. Adv.* **2018**, *4* (4), eaaq1566. https://doi.org/10.1126/sciadv.aaq1566.